\documentclass[aps,pre,reprint,amsmath,amssymb,showpacs,superscriptaddress]{revtex4-2}
\usepackage{graphicx}% Include figure files
\usepackage{dcolumn}% Align table columns on the decimal point
\usepackage{bm}% bold math
\usepackage{color}
\usepackage{lipsum}
\usepackage{hyperref}% add hypertext capabilities
\usepackage{mathrsfs} 
\usepackage{lineno}
\usepackage{siunitx}
\usepackage{multirow}

\hypersetup{colorlinks=true,
            linkcolor=blue,
            anchorcolor=blue,
            citecolor=blue,
			urlcolor=blue}

\begin{document}

\title{Structure of weakly collisional shock waves of multicomponent plasmas inside hohlraums of indirect inertial confinement fusions}% Force line breaks with \\

\author{Tianyi Liang}
	\affiliation{Institute for Fusion Theory and Simulation, School of Physics, Zhejiang University, Hangzhou 310058, China}
\author{Dong Wu}
    \email{dwu.phys@sjtu.edu.cn}
    \affiliation{Key Laboratory for Laser Plasmas and School of Physics and Astronomy, and Collaborative Innovation Center of IFSA, Shanghai Jiao Tong University, Shanghai 200240, China}
\author{Lifeng Wang}
	\affiliation{Institute of Applied Physics and Computational Mathematics, Beijing 100094, China}
\author{Lianqiang Shan}
	\affiliation{National Key Laboratory of Plasma Physics, Research Center of Laser Fusion, CAEP, Mianyang 621900, China}
\author{Zongqiang Yuan}
	\affiliation{National Key Laboratory of Plasma Physics, Research Center of Laser Fusion, CAEP, Mianyang 621900, China}
\author{Hongbo Cai}
	\affiliation{Institute of Applied Physics and Computational Mathematics, Beijing 100094, China}
\author{Yuqiu Gu}
	\affiliation{National Key Laboratory of Plasma Physics, Research Center of Laser Fusion, CAEP, Mianyang 621900, China}
\author{Zhengmao Sheng}
	\email{zmsheng@zju.edu.cn}
	\affiliation{Institute for Fusion Theory and Simulation, School of Physics, Zhejiang University, Hangzhou 310058, China}
\author{Xiantu He}
	\affiliation{Institute for Fusion Theory and Simulation, School of Physics, Zhejiang University, Hangzhou 310058, China}
	

\begin{abstract}

	In laser-driven indirect inertial confinement fusion (ICF), a hohlraum--a cavity constructed from high-Z materials--serves the purpose of converting laser energy into thermal x-ray energy. This process involves the interaction of low-density ablated plasmas, which can give rise to weakly collisional shock waves characterized by a Knudsen number $K_n$ on the order of 1. The Knudsen number serves as a metric for assessing the relative importance of collisional interactions. Preliminary experimental investigations and computational simulations have demonstrated that the kinetic effects associated with weakly collisional shock waves significantly impact the efficiency of the implosion process. Therefore, a comprehensive understanding of the physics underlying weakly collisional shock waves is essential. This research aims to explore the formation and fundamental structural properties of weakly collisional shock waves within a hohlraum, as well as the phenomena of ion mixing and ion separation in multicomponent plasmas. Weakly collisional shocks occupy a transition regime between collisional shock waves ($K_n \ll 1$) and collisionless shock waves ($K_n \gg 1$), thereby exhibiting both kinetic effects and hydrodynamic behavior. These shock waves are primarily governed by an electrostatic field, which facilitates significant electrostatic sheath acceleration and ion reflection acceleration. The differentiation of ions occurs due to the varying charge-to-mass ratios of different ion species in the presence of electrostatic field, resulting in the separation of ion densities, velocities, temperatures and concentrations. The presence of weakly collisional shock waves within the hohlraum is expected to affect the transition of laser energy and the overall efficiency of the implosion process. 

\end{abstract}
\maketitle
%\table of contents   

\section{\label{sec:intro}INTRODUCTION}

\begin{figure}[htbp]
		\centering
		\includegraphics[scale=0.33]{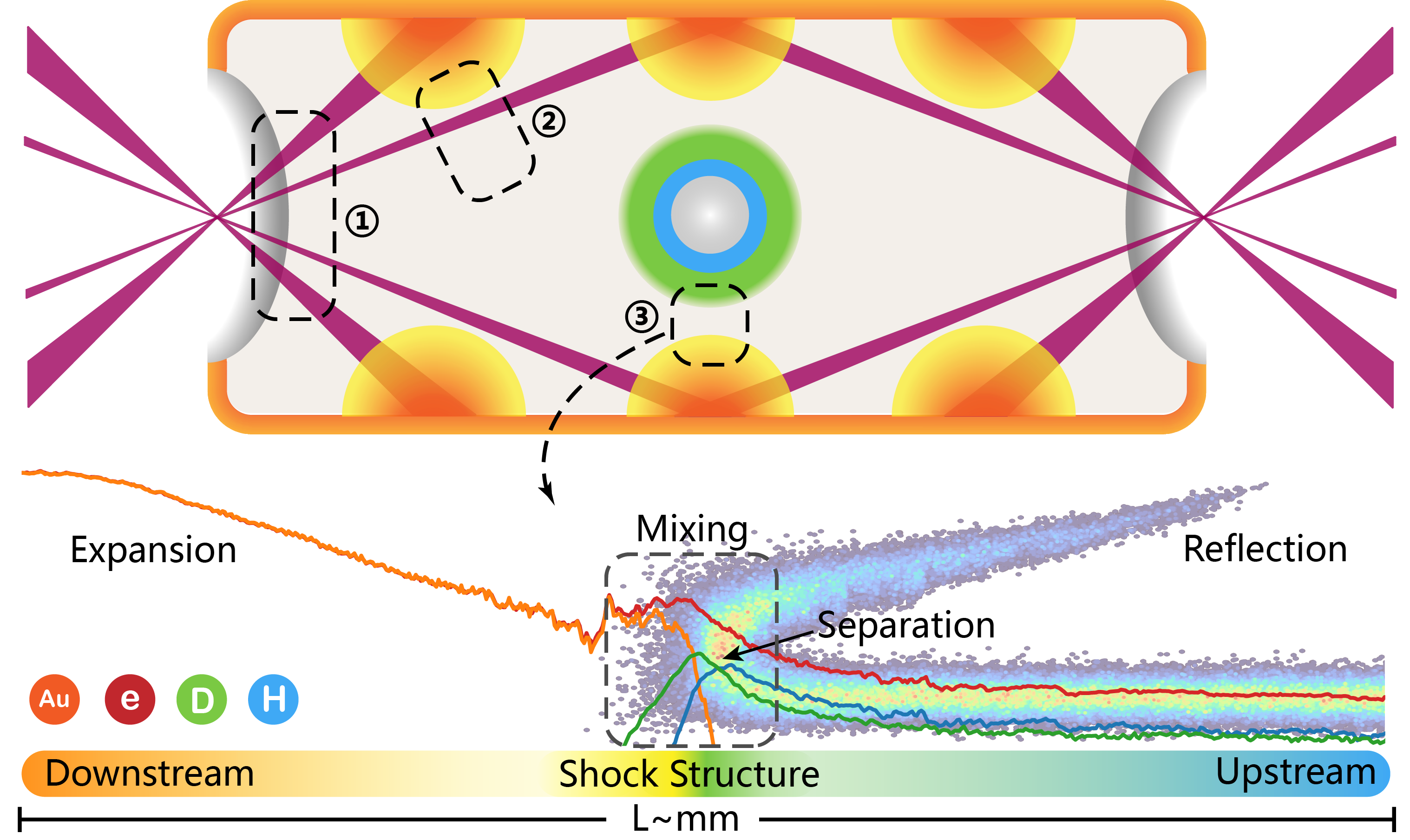}
		\caption{The schematic diagram of three kinds of interaction regions for the colliding plasmas within a hohlraum of laser-driven indirect ICF. Region 1 pertains to the interaction between the sealing films and filled gas. Region 2 focuses on the interaction between gold (Au) bubbles and filled gas. Region 3 addresses the interaction between Au bubbles and the ablated fusion fuel plasmas, such as the hydrogen-deuterium (HD) plasmas. An enlarged image of Region 3 illustrates various phenomena, including plasma expansion, the formation of weakly collisional shock, collisionless ion reflection, ion mixing and ion separation.}
		\label{fig:fig1} 
	\end{figure}

	In the context of laser-driven indirect inertial confinement fusion (ICF), the laser energy is transformed into the thermal x-ray energy within a cavity known as a hohlraum. This hohlraum is constructed from high atomic number (Z) materials, particularly gold. This process generates a high ablation pressure on the surface of a fuel-containing capsule as well as on the inner wall of the hohlraum, leading to the formation of low-density ablated plasmas. Within the framework of plasma interactions occurring outside the capsule in a gas-filled or near-vacuum hohlraum \cite{Pape2016}, three corresponding regions can be identified, as indicated by radiation hydrodynamics simulations \cite{Hopkins2015prl,Xie2022,Yan2019}. The schematic diagram of these regions are illustrated in Fig.~\ref{fig:fig1}. 
	
	The first region involves the interactions between sealing films and the gas, typically helium (He), that occupies the hohlraum, where collisional effects are minimal. These sealing films are designed to contain the gas, and are initially subjected to laser ablation, subsequently becoming entrained within the gas. The second region involves the interaction between gold bubbles generated from laser ablation and the filled gas \cite{Thoma2017}. The third region encompasses the interactions between these gold bubbles and the plasmas produced from the fusion fuels as a result of X-ray ablation. All three regions are characterized by weak collisional effects owing to the low plasma density. Additionally, weakly collisional shock waves are likely to be present in these regions, which may influence the efficiency of the implosion process.

	A shock wave represents a notebale category of nonlinear propagating disturbance that travels at velocities exceeding the speed of sound within a specific medium. This phenomenon is characterized by abrupt variations in pressure, temperature, and density, among other properties. In contrast to conventional fluid media, plasma is composed of electrons and ions, resulting in a distinct shock wave structure. The structures strongly depend on the effect of collisions, which can be characterized by the Knudsen number. The Knudsen number defined as $K_n=\lambda\nabla\ln n$, serves as a dimensionless parameter that quantifies the relative significance of the mean free path (MFP) in relation to a characteristic length scale of the plasma, thereby providing an approximate measure of the influence of collisions within the plasma medium.

	Based on the Knudsen number the shock waves can be systematically categotized into four classifications: strongly collisional shock waves (characterized by $K_n \ll 1$, typically $K_n \le 10^{-3}$), moderately collisional shock waves ($K_n\sim 0.01-0.1$), weakly collisional shock waves (with moderately large $K_n\sim \mathcal{O}(1)$) and collisionless shock waves ($K_n \gg 1$). As the Knudsen number $K_n$ increases, the significance of kinetic effects becomes increasingly pronounced \cite{Simakov2017}. In the context of collisionless shock waves, kinetic effects predominantly govern their structural characteristics. Moreover, even in moderately collisional shock waves, it is essential to consider the impact of kinetic effects \cite{Keenan2018}. Weakly collisional shocks ($K_n \sim \mathcal{O}(1)$) are in the transition region between collisional shock waves ($K_n \ll 1$) and collisionless shock waves ($K_n \gg 1$). In this regime, both kinetic effects and hydrodynamic behavior are present \cite{Zhang2024}, with the predominant influence being contingent upon the specific value of the Knudsen number. Consequently, it is imperative to elucidate the characteristics of both collisional and collisionless shock waves.
	
	The structure of strongly collisional shocks within plasmas has been extensively investigated \cite{Jukes1957, Greywall1975}. Preliminary hydrodynamic estimates, which incorporate factors such as electron heat conduction, indicate that the temperatures of electrons and ions beomes decoupled at the shock front \cite{zeldovich}. Consequently, the shock front can be delineated into three distinct regions: the pre-heating layer, the shock region and the equilibration layer \cite{Keenan2017}. In a multicomponent plasma, the concentration of ion mass across the shock front is subject to perturbation. For collisional shock waves, these characteristics can be effectively represented by the multi-ion hydrodynamic framework, specifically the Braginskii-Simakov-Molvig (BSM) model \cite{Simakov2016,Simakov2016-2,Simakov2017}, under conditions where the Knudsen number $K_n\ll 1$. 

	The issue of component separation resulting from collisional shocks has long been extensively investigated \cite{Cowling1942,Simakov2017,Keenan2018}, particularly in the context of laser-driven indirect ICF \cite{Rosenberg2014,Sio2019,Sio2019prl}. Indirect ICF process encompasses varying Knudsen numbers of shock waves and presents multicomponent problems. Within the capsule, strongly collisional shock waves compress the fuel toward the center, faciliating the conditions necessary for ignition, a process referred to as implosion. The inhomogeneity of the components can significantly diminish the energy conversion efficiency of the implosion \cite{Amendt2010,Yin2019}. Due to effective control over the implosion process, the National Ignition Facility (NIF) has successfully achieved ignition \cite{Zylstra2021,Zylstra2021prl,Zylstra2022,Zylstra2022pre,Atzeni2022}.

	Collisionless shock waves ($K_n \gg 1$) represent fascinating phenomena that manifest in a variety astrophysical and space environments \cite{Balogh2013}. These environments include supernova remnants \cite{Spicer1990}, solar flares \cite{Burlaga2008}, and the Earth's magnetosphere \cite{Sagdeev1991, Bertucci2005}. It is posited that collisionless shocks play a significant role in the generation and amplification of magnetic fields \cite{Gregori2015}, as well as in the acceleration of particles to energies that are characteristic of cosmic rays throughout the universe \cite{Colgate1960}. Given that the MFPs are considerably larger than the characteristic lengths involved, dissipation effects cannot be attributed to collisions; instead, alternative dissipation processes, such as ion acceleration, are expected to compensate for this \cite{Romagnani2008,Sorasio2006}. Consequently, the structural characteristics of collisionless shock waves are inherently more complex and distinct from those of collisional shocks. Furthermore, the charge-to-mass ratio of various ions is a critical factor that influences these kinetic processes \cite{Kumar2021}. 
	
	For weakly collisional shock waves ($K_n \sim \mathcal{O}(1)$) in the Region 3 which is illustrated in Fig.~\ref{fig:fig1}, preliminary experimental investigations \cite{Shan2018} and computational simulations \cite{Cai2020,Liang2024,Zhang2024} have confirmed that their kinetic effects significantly impact the efficiency of the implosion process. For instance, certain energetic ions may be accelerated by an electrostatic shock \cite{Ahmed2013,Rinderknecht2018}, subsequently entering the capsule and resulting in an abnormal broader spectrum of neutron production. The intricacy of ion mixing within this region depends on the design of the ablator utilized in the capsule. Specifically, in the case of plastic (CH) or high density carbon (HDC) materials \cite{Pape2014,Hopkins2015}, the region involves gold (Au), carbon (C) ions and hydrogen isotopes. It is noteworthy that these kinetic effects, along with ion separation and mixing phenomena cannot be accurately simulated using radiation hydrodynamics codes. Instead, a purely kinetic code or a hybrid-kinetic code \cite{Cai2021} is required for accurate representation.
 
	This study investigates the formation and fundamental structure of a weakly collisional shock wave generated from the collision between a gold plasma and a multicomponent plasma within a hohlraum. Additionally, it explores the phenomena of ion separation and mixing across the shock wave. Utilizing a high-order implicit Particle-In-Cell (PIC) code, referred to as LAPINS \cite{Wu2017,Wu2019,Wu2021,Wu2020,Wu2023}, we conduct large-scale kinetic simulations to replicate the entire process. To elucidate the properties of the shock waves, a comparation has been conducted between weakly collisional, strongly collisional, and collisionless shock waves. What's more, the effect of varying mole-fraction ratios has been examined.

	The paper is organized as follows. In multicomponent plasmas, the overall structure and nature of the shock wave is discussed in Sec.~\ref{sec:sec2}, including the setup of simulations. In Sec.~\ref{sec:sec3}, the separation and mixing of ions with different charge-to-mass ratios in shock waves are discussed, as well as the effect of different mole-fraction ratios on the shock wave structure. Finally, the conclusion are displayed in Sec.~\ref{sec:conclusion}.
 
\section{\label{sec:sec2}weak collisional shock in colliding multicomponent plasmas} 
      
    The one-dimensional implicit PIC version of the LAPINS code is utilized for the simulations, which examines the interaction between an expanding gold plasma and a hydrogen-deuterium (HD) plasma, as depicted in Fig.~\ref{fig:fig1} (b). The simulation domain is defined with a length of $L=3\:\rm{mm}$, discretized into $3000$ computational cells. Each cell is populated with $1000$ macro-particles representing electrons and $400$ macro-particles corresponding to various ions. The gold plasma is positioned on the left side of the simulation domain, with a fixed ionization state of gold ions set at $Z=50$ and an initial ion temperature of $T_\text{Au}=100\:\rm{eV}$. The initial electron temperature of the gold plasma is $T_{e1}=3000\:\rm{eV}$, and the electron density is given by $n_\text{e1}=Zn_\text{Au}=1.0 \times 10^{21}\:\rm{cm^{-3}}$. On the right side of the simulation box, a full-ionized HD plasma with varying mole fractions, defined as $f_\text{H}=n_\text{H}/(n_\text{D}+n_\text{H})$, is present, characterized by an electron number density of $n_{e0}=2.0\times 10^{19}\:\rm{cm^{-3}}$, a proton number density of $n_\text{H}=f_\text{H}n_\text{e0}$ and a temperature of $T_\text{H}=T_\text{D}=T_\text{e0}=100 \: \rm{eV}$. The interaction between the expanding gold plasma and the HD plasma generates an electrostatic shock structure. The gold plasma is typically identified as the downstream region located behind the shock front, while the deuterium plasma is ragarded as the upstream region situated ahead of the shock. The sound speeds for the different ionic components within HD plasma are calculated as $c_\text{sH} = \sqrt{{kT_\text{e0}}/{m_\text{H}}}=113.0\:\rm{km/s}$ and $c_\text{sD}=79.8\:\rm{km/s}$. 

\begin{figure}[htbp] 
		\centering
		\includegraphics[scale=0.48]{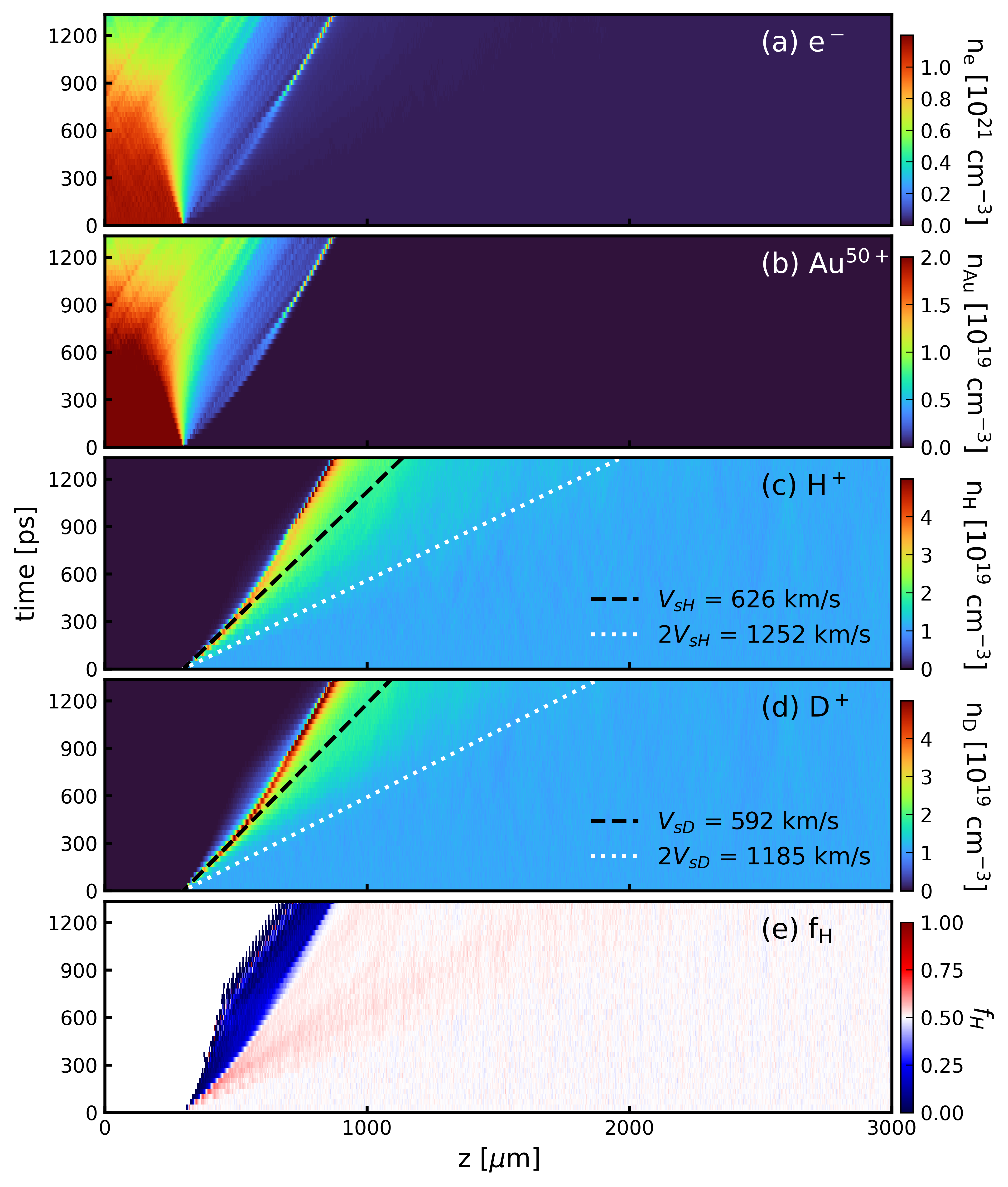}
		\caption{The temporal evolution of the density profiles of different species and mole-fraction $f_H$. (a) - (d) The temporal evolution of the density profiles of electrons ($\rm{e^-}$), and gold ions ($\rm{Au^{50+}}$), protons ($\rm{H^+}$), deuterium ions ($\rm{D^+}$) respectively. The black dashed lines in (c) and (d) depict the trajectory of the shock front, with a velocity of $V_\text{shH}=626\:\rm{km/s}$ in the proton component and $V_\text{shH}=592\:\rm{km/s}$ in the deuterium ion component, respectively. The white dotted lines depict twice of the velocity of the shocks. (e) The temporal evolution of mixing mole-fraction $f_\text{H}=n_\text{H}/(n_\text{H}+n_\text{D})$.}
		\label{fig:fig2} 
	\end{figure}

\begin{figure}[htbp]
		\centering
		\includegraphics[scale=0.45]{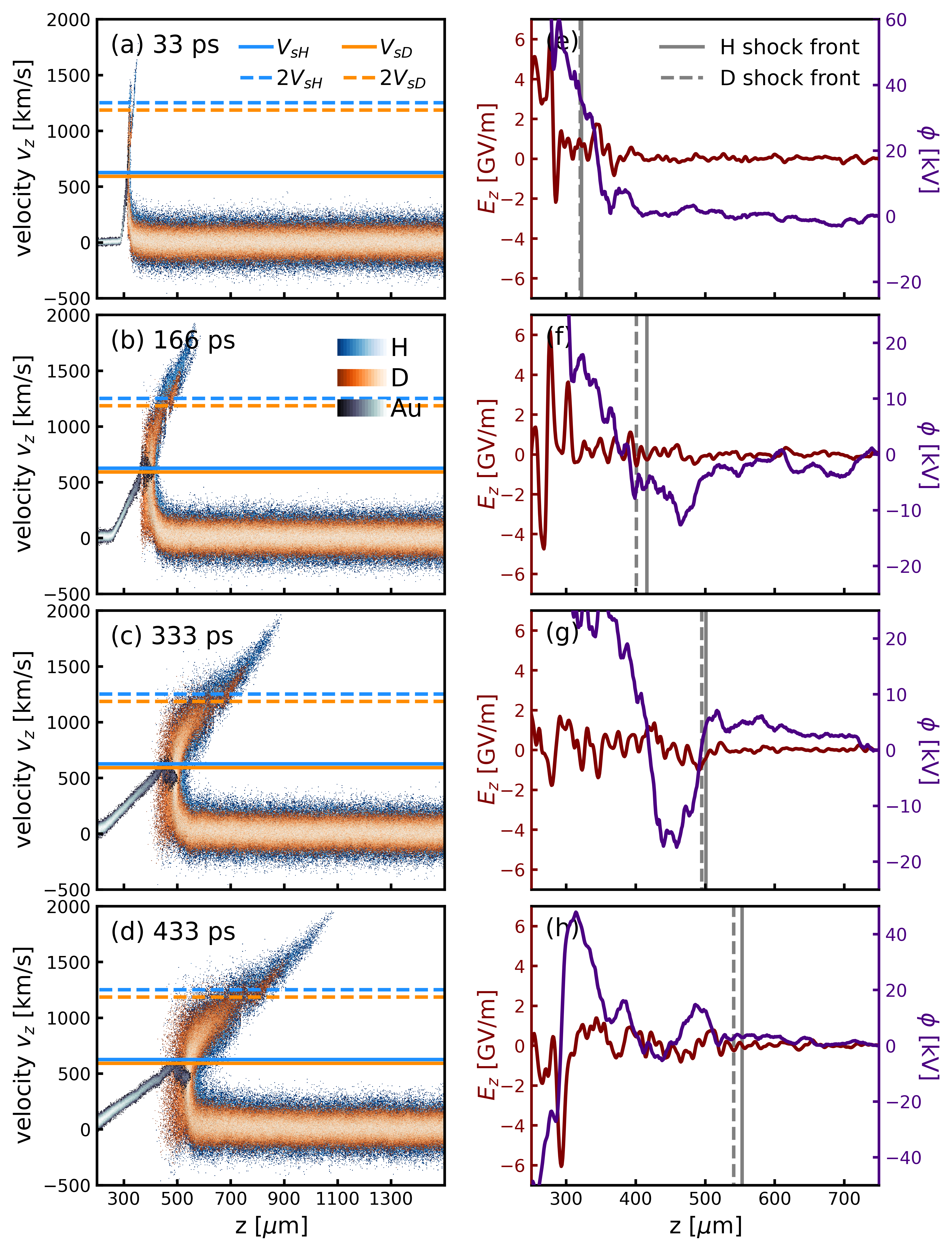}
		\caption{(a) - (d) The phase space distributions for protons ($\rm{H^{+}}$) and deuterium ions ($\rm{D^+}$) with initial mixing mole-fraction $f_\text{H}=0.5$ at the time of 33, 166, 333, 433 ps. The colormaps featuring blue, orange and gray are employed to represent the phase density of protons, deuterium ions and gold ions, respectively. The solid and dashed lines in blue and orange represent the velocity of the shock wave and its corresponding twice velocities of shock waves for hydrogen and deuterium ions, respectively. (e) - (h) The electric field, $E_z$ (left axis, red line) and potential, $\phi$ (right axis, purple line) at the corresponding times. The isopotential surface was selected at $\rm z = 750 \:\mu m$. The solid and dashed lines in gray are the positions of the shock front in the hydrogen and deuterium ions, respectively.}
		\label{fig:fig3}
	\end{figure}

	The entire physical process is elucidated through a simulation case with a fraction of $f_{\text{H}} = 1/2$, indicating that the densities of hydrogen and deuterium are equal, $n_\text{H}=n_\text{D}$. The temporal evolution of the density profiles for each species, along with the profile of $f_{\text{H}}$, is illustrated in Fig.~\ref{fig:fig2}. At the onset of the simulation, the higher-energy electrons, which process greater velocities than the gold ions, lead to the formation of an electrostatic sheath field characterized by charge separation. This electrostatic field initates a rarefaction expansion process within the gold plasma, rapidly accelerating both hydrogen and deuterium ions upstream to significantly high velocities, as shown in Fig.~\ref{fig:fig3} (a). An electrostatic shock wave is generated within the homogeneous HD plasma, driven by the expansion of the gold plasma. We quantified the velocities of the shock waves during the initial 300 picoseconds, noting a distinction between the velocities of hydrogen and deuterium ions. As illustrated in Fig.~\ref{fig:fig2} (c) and (d), the shock velocities for hydrogen and deuterium ions are measured at $V_{\text{sH}}=626\:{\rm km/s}$ and $V_{\text{sD}}=592\:{\rm km/s}$, respectively. Beyond the 300 picosecond mark, a significant reduction in the shock    velocity, $V_{\text{sH}}$, for hydrogen ions is observed, as indicated by the steepening of the red segment of the density distribution. In contrast, the decrease in $V_{\text{sD}}$ for deuterium ions is comparatively less pronounced. 
	
	In Fig.~\ref{fig:fig2} (e), the depicted area can be approximately divided into two distinct regions: the blue area, which is located downstream, and the light-red area, which is found in the foreshock region. The difference in mass between hydrogen and deuterium ions results in a phenomenon whereby the lighter hydrogen ions are more likely to experience acceleration upon reflection into the upstream region, thus contributing to the formation of the foreshock region. In contrast, the heavier deuterium ions in the downstream region are more likely to be deposited following the shock wave, leading to the establishment of the blue region. Within this region, the collision frequencies between gold ions and hydrogen or deuterium ions play a crucial role. These hydrogen and deuterium ions can be treated as test particles, and the collision frequency between them and gold ions is expressed as $\nu_{\text{i, Au}} = 4/3\pi^{1/2}({Z_\text{Au}Z_\text{D}}/{4\pi\varepsilon_0})^2n_\text{Au}{m_\text{i}^{-1/2}E_\text{i}^{-3/2}}\ln\Lambda$, where the subscript $i$ represents either $\rm H$ or $\rm D$. Consequently, the relationship between the collision frequencies can be described as $\nu_{\text{H, Au}}/\nu_{\text{D, Au}}\sim \sqrt{2} (E_D/E_H)^{3/2}$, with $\nu_{\text{D, Au}} \sim 10^{12}-10^{13}/s$. Therefore, in the shock region, $K_n\sim\mathcal{O}(1)$, indicating that the MFPs and the characteristic lengths of the shock wave are comparable. This suggests that the kinetic effects are significant, while the collision effects partially counterbalance the kinetic effects.

	The phase space distribution of hydrogen and deuterium ions at various time intervals reveals a clear kinetic effect, characterized by the consistent acceleration and reflection of both ion species. These ions experience identical acceleration due to the electrostatic sheath field and are reflected by the electrostatic shock wave. As illustrated in Fig.~\ref{fig:fig2}, the speeds of the shock waves differ between hydrogen and deuterium ions. The representation of these ions within the phase space [Fig.~\ref{fig:fig3} (a) - (d)] enhances the analysis of the impact of electrostatic shock wave reflection on hydrogen and deuterium ions denoted by the green and blue solid lines. The phase space at various moments indicates the development of weakly collisional shock waves and the formation of a C-shaped structure resulting from the reflection of upstream ions. The velocities of both shock waves are located at the center of this C-shaped structure. The velocity plateau, which arises from the steady reflection of the upstream ions by the shock, is clearly depicted in Fig.~\ref{fig:fig3} (d). The orange dashed line representing $2V_\text{sD}$ is positioned nearly at the midpoint of the velocity plateau for the deuterium ions, while the velocity plateau for the hydrogen ions is slightly below the blue dashed line indicating $2V_\text{sH}$. This discrepancy is attributed to the gradual decrease in shock wave velocities for hydrogen ions.

    The reflection of the collisionless shock wave on the upstream ions can be understood as the scattering of the ions by the potential barrier, denoted as $\Delta \phi$, when the velocity of the upstream ions, $v_i$, satisfies,

	\begin{equation}
		Z_i e\Delta \phi \geq 1/2m_i(v_i-V_{s})^2.
		\label{eq:eq1}
	\end{equation}

	Here, $V_{s}$ represents the shock velocity, while the subscript $i$ indicates different species of ions. As illustrated in Fig.~\ref{fig:fig3} (e)-(g), a pronounced potential barriers in the vicinity of the shock fronts (indicated gray lines). 

	\begin{figure}[htbp]
		\centering
		\includegraphics[scale=0.44]{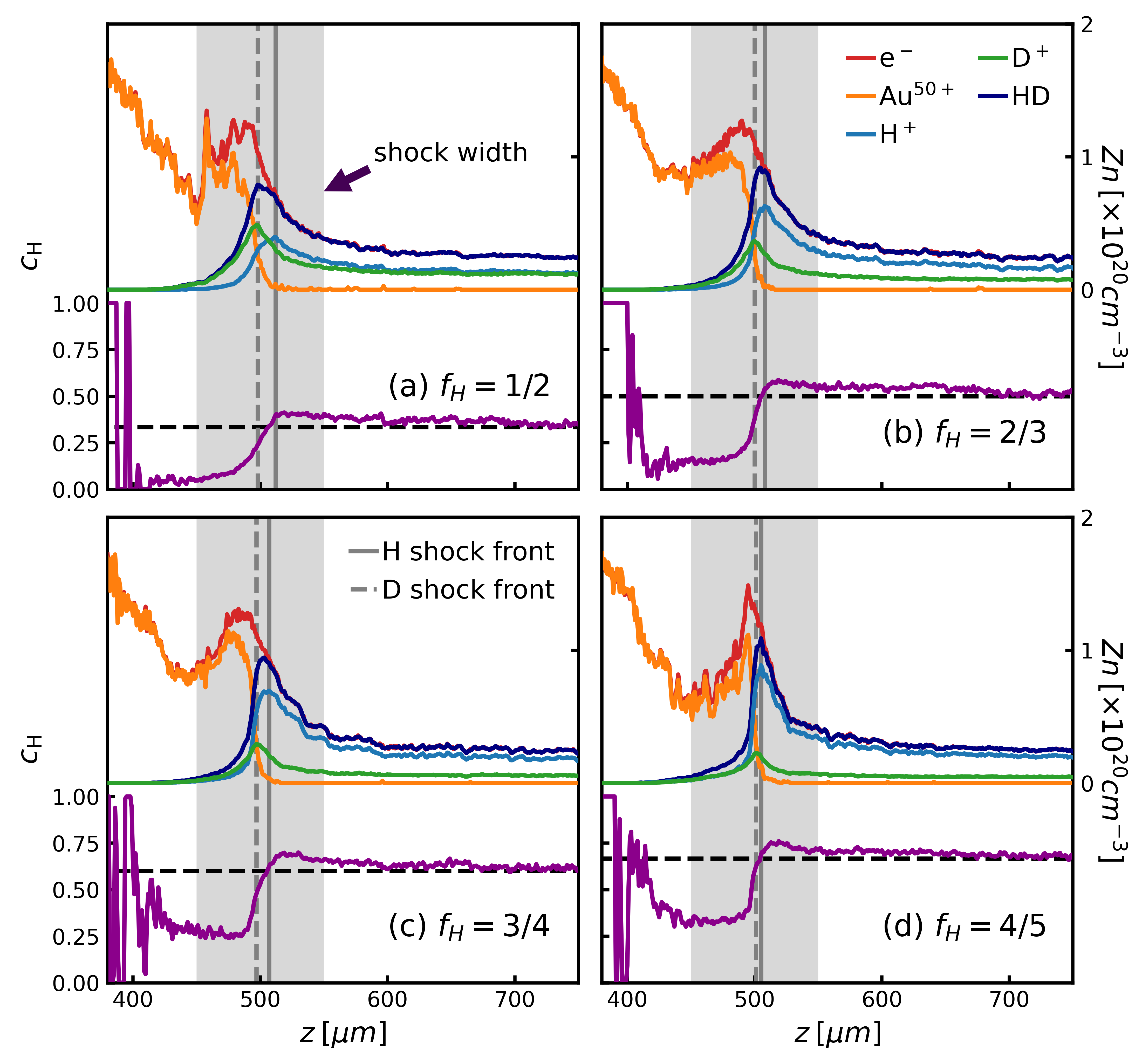}
		\caption{(a) The profile of flow velocity $u_z$ of hydrogen ions ($\rm H^+$) and deuterium ions ($\rm D^+$) at the time of 333 ps at different initial mole-fraction of $1/2$, $2/3$, $3/4$, and $4/5$. Solid lines in the red series and green series indicate hydrogen and deuterium ions, respectively. The gradual color change indicates the different fractions. The solid green and blue horizontal lines indicate the shock velocity among hydrogen ($V_\text{sH}$) and deuterium ions ($V_\text{sD}$) measured in Fig.~\ref{fig:fig2}, respectively. The gray-filled range in the background indicates the approximate width of the shock wave, and the gray solid and dashed lines are the positions of the shock front in the hydrogen and deuterium ions, respectively. The dark-gray-filled range indicates the pre-heat region of the shock wave. (b) The time-integrated velocity spectra ($v_z$) of hydrogen and deuterium ions with the diagnostic plane at $\rm z=650\: \mu m$. The blue and green dashed lines indicate twice the value of the shock velocity in hydrogen and deuterium ions, respectively. (c) Ion temperatures ($T_i$) of hydrogen and deuterium ions. (d) Energy spectra of hydrogen and deuterium ions.}
		\label{fig:fig4}
	\end{figure}

	At the diagnostic plane located at $\rm z=650\: \mu m$, the time-integrated velocity and energy spectra have been recorded, as illustrated in Fig.~\ref{fig:fig4} (b) and (d).  As depicted in Fig.~\ref{fig:fig3} (b), the maximum velocity attained by the hydrogen ions is approximately $2000\:\rm{km/s}$, whereas deuterium ions reach a maximum velocity around $1500\:\rm{km/s}$. These findings are in alignment with the observed velocity spectra. Both protons and deuterium ions exhibit favorable quasi-monoenergetics characteristics, which is indicative of shock wave reflection acceleration. Additionally, due to the presence of electrostatic sheath acceleration, the energy at the peak for both ions speeds exceeds $2V_{s}$.

	We calculated the average velocity of particle movement within each mesh and utilized this information to define the flow velocity of species, denoted as $u_{z,s}\equiv\sum v_{i,s}\omega_i/\sum\omega_i$, where $\omega_i$ represents the weight of macro-particles within a grid. As illustrated in Fig.~\ref{fig:fig4} (a), the flow velocity exhibits a single-peaked profile centered around the shock front. This phenomenon can be readily comprehended. The ions in the upstream region are reflected and accelerated by the shock. Consequently, in proximity to the shock front, a significant proportion of ions attain elevated velocities. As one move further from the shock front, this proportion diminishes gradually, and the ions experience energy loss due to weak collisions. The resultant average indicates a progressive decline in flow velocity. Notably, the flow velocities of both ion species are lower than the velocity of shock wave. Therefore, the characteristics of a weak collision shock wave differ markedly from those of a collisional shock wave.

	The temperature of a species can be expressed as $T_s\equiv 1/2m_s\sum(v_{i,s}-u_{z,s})^2\omega_i/\sum\omega_i$. This formulation is predicated on the assumption that the particles within the grid achieve thermal equilibrium. In scenarios where this assumption does not hold, it is more appropriate to refer to this quantity as thermal kinetic energy. Within the pre-heating layer, the temperatures of both ions gradually increase, following by a sudden decrease at the shock front. The behavior diverges from that observed in conventional collisional shock waves \cite{zeldovich}. 
	
	As previously discussed, the upstream ions are reflected at the shock front, which can be interpreted as a transition from disorder to order. Consequently, the temperature, which characterizes the irregular motion of the ions, experiences a sharp decline, although it remains above the initial temperature. Upon entering the downstream region, the temperature of ions continues to rise. According to hydrodynamic theory, for strongly collisional shock waves, the downstream ion temperature can be approximated as $T_{ic}\sim 1/2m_iV_{s}^2$ \cite{zeldovich}. Therefore, the expected downstream temperature for both ions species are $T_\text{H,1}\sim 1.8\:\rm{keV}$ and $T_\text{D,1}\sim 3.6\:\rm{keV}$, while the temperatures depicted in Fig.~\ref{fig:fig4} (c) are approximately $T_i\sim2.1\:\rm{keV}$. Although the temperatures obtained may lack statistical significance due to the limited number of ions present downstream, they nonetheless provide a rough indication of the high energy accessible to ions traversing through the shock front.

	\section{\label{sec:sec3}Ion separation and mixing at weakly collisional shock}

	In the preceding section, we discussed the fundamental characteristics of weak collisional shock waves ($K_n\sim\mathcal{O}(1)$) that arise from plasma collision processes within the hohlraum environment. In scenarios where the background plasma consists of multiple components, the varying charge-to-mass ratios of the ions result in the phenomena of separation and mixing among the ionic components, which subsequently alters the structure of the shock wave. For the purposes of this discussion, we define the ion separation as the alteration of physical properties, including velocity, temperature, and mole-fraction of different ion species, within the initially homogeneous background plasma as a consequence of the shock wave's influence. Conversely, ion mixing refers to the interpenetration of different plasma components during the colliding process.

	\subsection{\label{sec:subv} Velocity and Temperature Separation}

	A more intricate structure is observed for collisionless shock waves as they traverse a multicomponent plasma. In the absence of collisions among ions, each ion speices independently develops its own sub-shocks in the presence of an electromagnetic field, concurrently encountering multiple potential barriers \cite{Kumar2021}. This phenomenon is also evident during the initial phases of weakly collisional shock wave formation. As illustrated in Fig.~\ref{fig:fig2} (c) and (d), the shock velocities for difference ion speices exhibit variability, with a discernible gap of approximately $30\:\rm{km/s}$. The range of shock velocities can be obtained from Eq.~(\ref{eq:eq1}),
	\begin{equation}
		v_i + \sqrt{\frac{2e\Delta\phi}{m_i}} \geq V_{sh} \geq v_i - \sqrt{\frac{2e\Delta\phi}{m_i}}.
		\label{eq:eq2}
	\end{equation}
	By substituting $v_i$ with the thermal velocity, $\sqrt{2kT_i/m_i}$, as indicated in Eq.~\ref{eq:eq2}, the ratio of shock velocities in the hydrogen and deuterium ions can be estimated as $V_\text{sH}/V_\text{sD}\sim\sqrt{m_\text{D}/m_\text{H}}=\sqrt{2}$. As measured in Fig.~\ref{fig:fig2},the observed ratio is $V_\text{sH}/V_\text{sD}=1.06$. Consequently, in the context of weakly collisional shock waves, the velocity separation of sub-shocks is less pronounced compared to the collisionless xounterparts. On the other hand, as the energy of the shock wave dissipates, the velocities of the two sub-shock waves gradually diminish and converge towards a common value.

	The disparity in mass and initial velocities between the hydrogen ions and deuterium ions results in the former achieving higher velocities over time. This phenomenon can be attributed to the fact that the charge-to-mass ratio of hydrogen ions exceeds that of the deuterium ions, allowing hydrogen ions to be the first to respond to the  electrostatic sheath field. As hydrogen ions accelerate and approach the electrons, they effectively shield and diminish the strength of the electrostatic sheath field, thereby reducing its impact on the deuterium ions. 
	
	Consequently, the maximum velocity attained by hydrogen ions suspasses that of the deuterium ions, and the population of hydrogen ions with high velocity is greater than that of deuterium ions. The ratio of maximum velocities is approximately $v_\text{H,max}/v_\text{D,max}\thickapprox 1.33$. As a result, in the energy spectra, the maximum energy of deuterium ions is observed to be greater than that of hydrogen ions. If the resolution of the experimental diagnostic energy spectra is adequate, it should be feasible to detect two distinct peaks in ion energy spectra, which correspond to the hydrogen and deuterium ions, respectively.
	
	Furthermore, there exist differences in the flow velocity and temperature between hydrogen and deuterium ions. The flow velocity of hydrogen ions is marginally greater than that of deuterium ions, attributable to the highter velocities of accelerated hydrogen ions compared to the deuterium ions. The temperature profiles for deuterium ions exhibit significant divergence from those of hydrogen ions. Initially, hydrogen ions experience heating; however, deuterium ions ultimately attain a higher temperature. Additionally, deuterium ions display a minor heating peak in the pre-heating layer, which is likely a consequence of energy deposition from reflected ions in that region. Both ion species ultimately achieve the same temperature behind the shock front. 

	\subsection{\label{sec:subc} Concertreation Separation}

	From total ion mass conservation equation, 
	\begin{equation}
		\label{eq:eq3}
		\frac{\partial \rho}{\partial t} + \nabla \cdot (\rho \bm{u}) = 0,
	\end{equation}
	the total ion mass flux remains constant through a steady-state shock, that is, 
	\begin{equation}
		\label{eq:eq4}
		\rho u=\rho_0 u_0.
	\end{equation}
	While from hydrogen mass conservation equation, 
	\begin{equation}
		\label{eq:eq5}
		\frac{\partial \rho_\text{H}}{\partial t} + \nabla \cdot [\rho_H (\bm{u}+\bm{v_\text{H}})]=0,
	\end{equation}
	the hydrogen mass flux is given as follow,
	\begin{align}
		\label{eq:eq6}  
		\rho u c_\text{H} + i = \rho_0 u_0 c_\text{H0}=0.
	\end{align}
	The diffusion flux of the hydrogen ions, $i_\text{H}$, can be expressed as follow \cite{LANDAU1987227,Kagan2012,Keenan2018},
	\begin{align}
		\bm{i_\text{H}} &\equiv \rho_\text{H}\bm{v_\text{H}} = \rho_\text{H}(\bm{u}_H-\bm{u}) \nonumber \\
		&= -\rho D\nabla c_\text{H} - \rho D \left( \frac{\kappa_p}{P_i}\nabla P_i\right) \nonumber \\
		&-\rho D \left( \frac{\kappa_{Ti}}{T_i}\nabla T_i + \frac{\kappa_{Te}}{T_e}\nabla T_e\right) \nonumber \\
		&-\rho D \left( \frac{e\kappa_E}{T_i}\nabla\phi\right).
		\label{eq:eq7}
	\end{align}

	Here, $\rho_\text{H}=n_\text{H}m_\text{H}$ represents the mass density of the hydrogen ions and $\rho=\rho_\text{H}+\rho_\text{D}$ denotes the total ion mass density. Concentration of hydrogen ion is denoted as $c_\text{H}\equiv\rho_\text{H}/\rho$, while the mass-averaged ion flow velocity is defined as $\bm{u} \equiv (\rho_\text{H}\bm{u}_\text{H}+\rho_\text{D}\bm{u}_\text{D})/\rho$. The subscript $0$ presents the quantities of plasma in the upstream region. 

	The driving forces contributing to the separation of HD plasma include the gradient of ion pressure ($\nabla P_i/P_i$), the gradient of ion temperature ($\nabla T_i/T_i$), and the electric field gradient ($e\nabla\phi/T_i$), which correspond to baro-diffusion, thermo-diffusion, and electro-diffusion, respectively. The cofficients, $D$, $\kappa_p$, $\kappa_{Ti}$, $\kappa_{Te}$and $\kappa_E$ can be articulated from the BSM models \cite{Simakov2016,Simakov2016-2,Simakov2017,Keenan2018}.
	
	From Eq.~\ref{eq:eq4} and Eq.~\ref{eq:eq6}, the variation in the concentration of the hydrogen ions is articulated as,
	\begin{equation}
		\label{eq:eq8}
		c_\text{H} - c_\text{H0} = -\frac{i}{\rho u}.
	\end{equation}
	Therefore, there are an enhancement of the hydrogen concertreation because of differential diffusion effects. In the context of strongly collisional shocks ($K_n \ll 1$), baro-diffusion becomes significant due to the pronounced pressure gradient at the shock front \cite{Amendt2010}. This baro-diffusion phenomenon results in an enhancement of hydrogen ion concentration at the shock front, accompanied by a corresponding deficit in the rearward region when the shock wave is nonstationary. Conversely, in collisionless simulations ($K_n\gg 1$), the effects of diffusion are predominantly influenced by the electric field. The presence of sub-shocks is more pronounced, and the potential barriers associated with each sub-shock can be distinctly observed through simulations \cite{Kumar2021}. The reflection and acceleration of ions significantly influence the structure of shock waves. Furthermore, electro-diffusion may play a crucial role in the separation of ion species due to the varying responses of particles with different charge-to-mass ratios to the electric field.  
	
	\begin{figure}[htbp]
		\centering
		\includegraphics[scale=0.45]{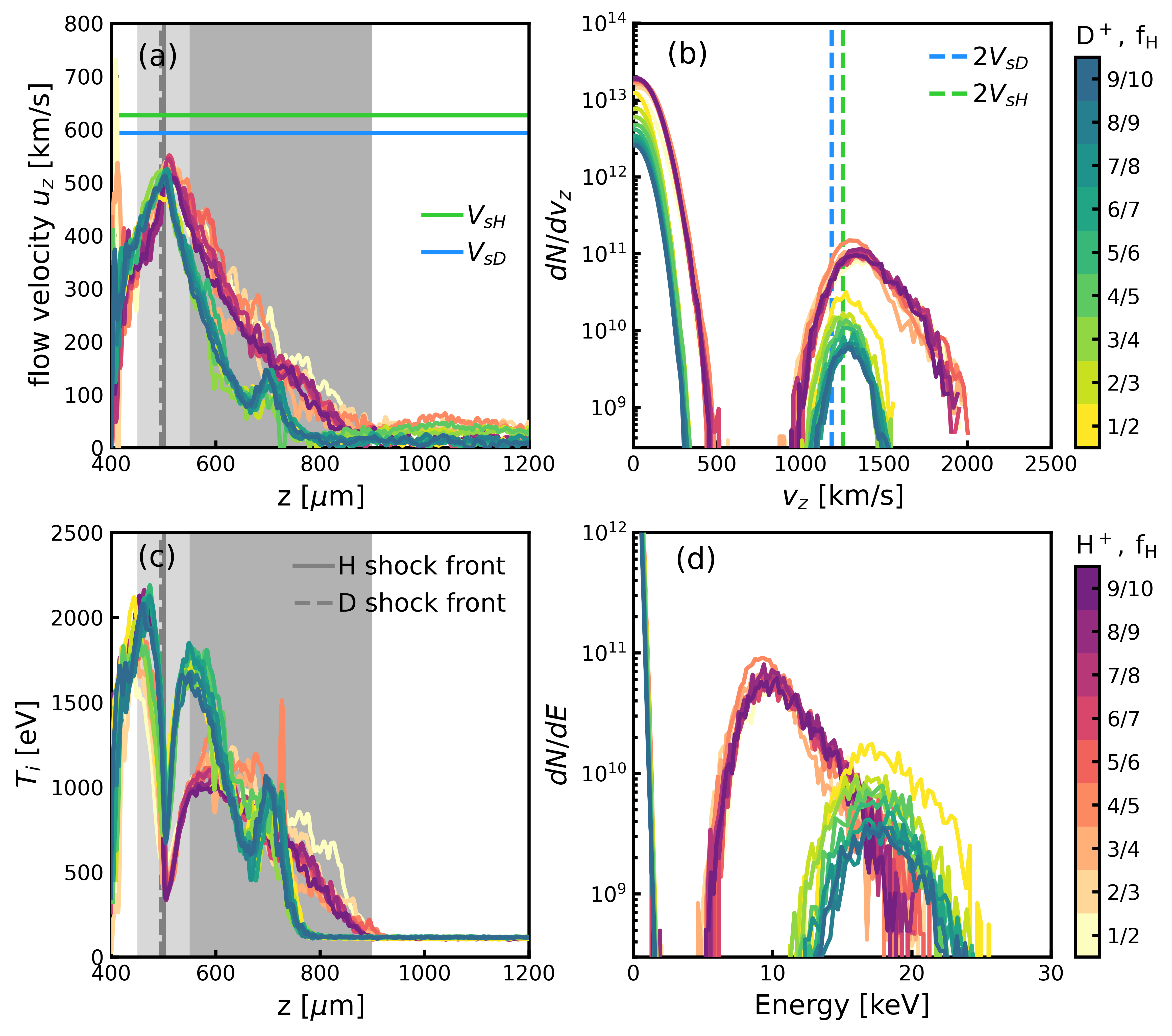}
		\caption{(a) - (d) The charge density profiles of various species, denoted as $Zn$ on the right axis, alongside the concentration profile of hydrogen ions denoted as $\rm f_\text{H}$ on the left axis at the time point of 333 ps. The initial mole-fractions are set as $1/2$, $2/3$, $3/4$, and $4/5$, respectively. The orange lines at the top correspond to gold ions ($\rm Au^{50+}$), while the red lines represent electrons ($\rm e^-$). The blue lines indicate the hydrogen ions ($\rm H^+$), the green lines denote the deuterium ions ($\rm D^+$), and the dark blue lines reflect the combined charge densities of hydrogen and deuterium ions ($\rm HD$). The lower purple lines depict the concentration profiles of hydrogen ions ($c_\text{H}$), with the black dashed lines representing the corresponding initial concentration value. The gray-filled region, indicated by an arrow in the background, approximates the width of the shock wave, while the gray solid and dashed lines delineate the position of the shock front for hydrogen and deuterium ions, respectively.}
		\label{fig:fig5}
	\end{figure}

\begin{figure}[htbp]
		\centering
		\includegraphics[scale=0.44]{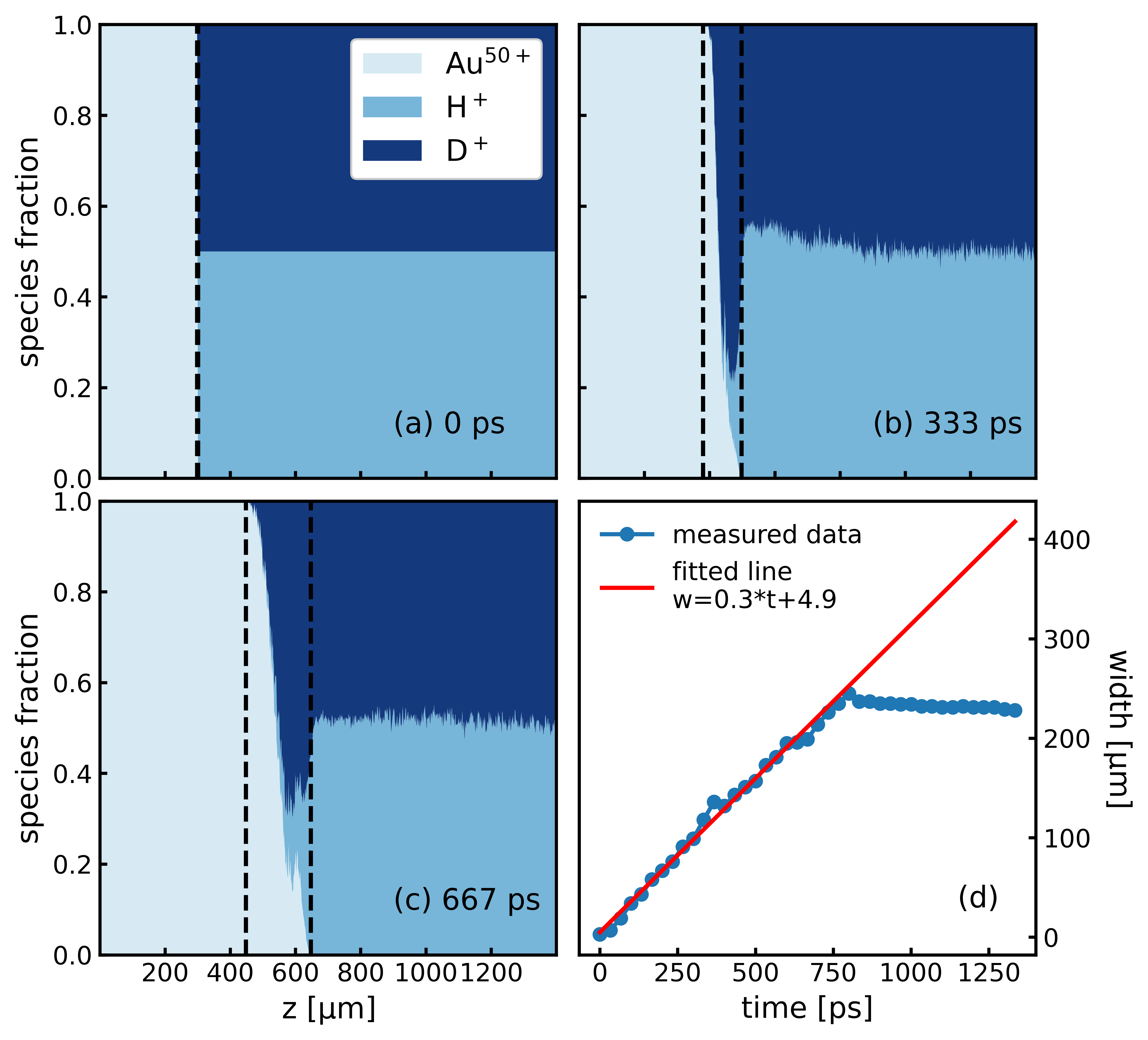}
		\caption{(a)-(c) The spatial distribution of species fractions at various time intervals (0 ps, 333 ps, and 667 ps). The areas represented in white blue, light blue, and dark blue correspond to the three ion species ($\rm Au^{50+}$, $\rm H^{+}$, and $\rm D^{+}$) respectively. The central region delineated by the two black dotted lines signifies the mixing region of the three ion species. The temporal variation in the width of this mixing region is illustrated in (d), with the red line representing the fitted curve.}
		\label{fig:fig6}
	\end{figure}

	These phenomena exhibit similarities to the outcomes observed in our weakly collisional simulations ($K_n\sim\mathcal{O}(1)$). In the density profiles, as depicted in Fig.~\ref{fig:fig6}, the separations between hydrogen and deuterium ions are more clearly represented. The peak of HD plasma can be divided into two sub-peaks of hydrogen ions and deuterium ions. Hydrogen ions seem to be pulled by electrons, while deuterium ions are pushed by gold ions. Specifically, the hydrogen ion concentration, $c_\text{H}$ is greater than $c_\text{H0}$ at the shock front and less than $c_\text{H0}$ in the downstream region, reflecting the nonstationary nature of the shock waves. However, the underlying mechanisms differ. Furthermore, in the presence of weak collisions, the kinetic effects dominated by the electric field have been suppressed.

	\subsection{\label{sec:subm} Ion Mixing} 

	In the absence of collisions, the expansion of gold plasma resembles free rarefaction rather than center rarefaction \cite{Atzeni2004}. The gold plasma undergoes a free expansion over time, characterized by a linear increase in its velocity. Conversely, when collisions are present, the expansion of the gold plasma is hindered, resulting in a linear increase in velocity that reaches a maximum value. In this scenario, the gold plasma functions analogously to a piston, faciliating the movement of the HD plasma, thereby enabling the mixing of the three ion species.
  
	To define the mixing width, we establish the left boundary at the position where $0.1\%$ of the hydrogen or deuterium ion density is found, and the right boundary at the position where $0.1\%$ of the gold ion density is located. The temporal variation of the mixing width is depicted in Fig.~\ref{fig:fig5} (d). The results from the linear regression indicate that the mixing width increases at a rate of $v_\text{mix}=0.3\:\rm{\mu m/ps}$ and then reaches saturation after expanding to $230\:\rm{\mu m}$. The phase of linear increase can be interpreted as indicative of the maximum relative velocity between the ions. The rate of increase in mixing width is contingent upon this maximum relative velocity. Due to their higher velocities, these ions exhibit longer MFPs and reduced collision frequencies. Consequently, in the shock rest frame the relative velocity can be expressed as $v_\text{mix} = v_\text{Au,max} + \max(v_\text{D, max},v_\text{H,max})$, where $v_\text{Au,max}$ represents the maximum speed at which the gold plasma expands to the right, while $v_\text{H,max}$ and $v_\text{D, max}$ denote the maximum velocity of diffusion of hydrogen and deuterium ions to the left, respectively. When these ions are sufficiently hindered, the mixing width saturates and then slowly decreases.

\subsection{\label{sec:sube} Effects of Mole-fraction on Shock Wave Structure}

	The simulations with varying mole-fractions are conducted to investigate their influence on the structure of weakly collisional shock waves. The charge density ($Zn$) profiles of different species and the mole-fraction profiles ($f_\text{H}$) at the time of 333 ps for various initial mole-fractions are shown in Fig.~\ref{fig:fig5}. As increase in the mole-fraction $f_{\text{H}}$ corresponds to a decrease in the percentage of deuterium ions, leading the simulation results to progressively align with those of pure hydrogen plasma \cite{Liang2024}. Notably, the peak of the HD plasma (represented by the dark blue line) become both narrower and taller as $f_\text{H}$ increases, indicating that the strength of the shock wave intensifies with rising $f_\text{H}$. The existence of multiple component effects reduces the intensity of the shock wave and broadens its width.

	The blue and orange lines denote the charge densities of $\rm H^+$ and $\rm D^+$, respectively. At a mole-fraction of $f_\text{H}=1/2$, the charge density profile of HD plasma distinctly reveals two sub-peaks corresponding to $\rm H^+$ and $\rm D^+$, which are spatially separated. This separation confirms that the sub-peak associated with $\rm H^+$ arises from the electrostatic field, while the sub-peak for $\rm D^+$ is generated by the interaction of the gold plasma. Furthermore, it is evident that the distance of between the two shock fronts of sub-peaks diminishes as $f_\text{H}$ increases. As the percentage of deuterium ions decreases, the frequency of collisions between gold ions and deuterium ions also reduces. Consequently, collisions between hydrogen ions and gold ions become more significant, impeding the expansion of the gold plasma.
	Thus the ion density separation is weakened.
	
	As depicted in Fig.~\ref{fig:fig4}, the statistical physical quantities of the ions demonstrate a notable consistency across various mole-fractions $f_\text{H}$. This observation implies that the variations in these statistical physical quantities are not affected by the initial mole fraction; rather they may be associated with the response of the charge-to-mass ratio to the electrostatic shock wave. This finding highlights the significance of the electric field in the context of weakly collisional shock waves ($K_n\sim\mathcal{O}(1)$).

\section{\label{sec:conclusion}CONCLUSION}

	In summary, this paper examines the formation and fundamental structural characteristics of weakly collisional shock waves ($K_n\sim\mathcal{O}(1)$) within hohlraums, as well as the phenomena of ion mixing and ion separation in multicomponent plasmas. The interaction between gold plasma and HD plasma facilitates the movement of gold ions, which act akin to a piston, thereby establishing the weakly collisional shock wave structure. These shock waves are primarily influenced by an electrostatic field and are characterized significant electrostatic sheath acceleration and ion reflection acceleration. While the kinetic effects observed are analogous to those in collisionless shock waves ($K_n \gg 1$), but they are somewhat mitigated by the presence of weak collisions, rendering them more representative of actual conditions. Additionally, the collision of the reversed ions with the upstream ions resulting of the upstream ions, leading to the formation of a preheated layer. 
	
	The differentiation of ions occurs as a result of varying charge-to-mass ratios of different species of ions, when subjected to the electrostatic field, leading to the separation of ion densities, velocities and temperatures. It has been observed that the separation of ion densities diminishes with an increase in mole fraction $f_\text{H}$. These phenomena exhibit similarities to the behavior of strongly collisional shock waves, which are influence by factors such as heat conduction and baro-diffusion. Furthermore, kinetic effects may play a significant role in certain scenarios, such as the reflection of ions that can result in beam-target nuclear fusion reactions, as well as the implications of ion mixing and separation on the laser coupling efficiency. Specifically, within the context of the hohlraum, our large-scale simulations indicate that the mixing width of colliding plasmas interaction is on the order of a hundred micrometres, warranting further investigations into its impact on radiation hydrodynamic simulations. 

	The findings of this research partially address the existing gap in the investigation of shock wave structures in scenarios characterized by weak collision conditions. This study enhances the comprehension of the formation process of shock waves and their structural characteristics across varying Knudsen number conditions. Such insights are valuable for both astrophysics and laboratory astrophysics, specifically ICF, facilitating further research and the assessment of modifications to current models in light of collision effects. 

\begin{acknowledgments}
	This work is supported by National Natural Science Foundation of China (Grants No. 12075204 and No. 12235014), Foundation of National Key Laboratory of Plasma Physics, the Strategic Priority Research Program of Chinese Academy of Sciences (Grant No. XDA250050500), and Shanghai Municipal Science and Technology Key Project (Grant No. 22JC1401500). Dong Wu thanks the sponsorship from Yangyang Development Fund.

\end{acknowledgments}
 
\nocite{*}

\bibliography{references}% Produces the bibliography via BibTeX.

\end{document}